\documentclass[preprint,onecolumn,aps,pre,eqsecnum,a4paper]{revtex4-1}
\usepackage[latin9]{inputenc}
\usepackage{color}
\definecolor{note_fontcolor}{rgb}{0.800781, 0.800781, 0.800781}
\usepackage[english]{babel}
\usepackage{amsmath}
\usepackage{amssymb}
\usepackage{graphicx}
\usepackage[unicode=true,
 bookmarks=false,
 breaklinks=true,pdfborder={0 0 0},backref=false,colorlinks=true]
 {hyperref}

\makeatletter

\newenvironment{lyxgreyedout}
  {\textcolor{note_fontcolor}\bgroup\ignorespaces}
  {\ignorespacesafterend\egroup}

\@ifundefined{textcolor}{}
{%
 \definecolor{BLACK}{gray}{0}
 \definecolor{WHITE}{gray}{1}
 \definecolor{RED}{rgb}{1,0,0}
 \definecolor{GREEN}{rgb}{0,1,0}
 \definecolor{BLUE}{rgb}{0,0,1}
 \definecolor{CYAN}{cmyk}{1,0,0,0}
 \definecolor{MAGENTA}{cmyk}{0,1,0,0}
 \definecolor{YELLOW}{cmyk}{0,0,1,0}
}
\numberwithin{equation}{section}
\numberwithin{figure}{section}
\numberwithin{table}{section}

\usepackage{graphicx}

\DeclareGraphicsExtensions{.png,.pdf,.eps}
\graphicspath{{.}}

\makeatother

\begin{document}


\title{Conditions for Lorentz-invariant superluminal information transfer
without signaling}

\author{Gerhard \surname{Grössing}\textsuperscript{}}

\email[Corresponding author: ]{ains@chello.at}

\homepage{http://www.nonlinearstudies.at}

\author{Siegfried \surname{Fussy}}

\author{Johannes \surname{Mesa Pascasio}}

\author{Herbert \surname{Schwabl}}

\affiliation{Austrian Institute for Nonlinear Studies, Akademiehof, Friedrichstr.~10,
1010 Vienna, Austria\vspace{1cm}
}
\begin{abstract}
We understand emergent quantum mechanics in the sense that quantum
mechanics describes processes of physical emergence relating an assumed
sub-quantum physics to macroscopic boundary conditions. The latter
can be shown to entail top-down causation, in addition to usual bottom-up
scenarios. With this example it is demonstrated that definitions of
``realism'' in the literature are simply too restrictive. A prevailing
manner to define realism in quantum mechanics is in terms of pre-determination
independent of the measurement. With our counter-example, which actually
is ubiquitous in emergent, or self-organizing, systems, we argue for
\emph{realism without pre-determination}. We refer to earlier results
of our group showing how the guiding equation of the de Broglie--Bohm
interpretation can be derived from a theory with classical ingredients
only. Essentially, this corresponds to a ``quantum mechanics without
wave functions'' in ordinary 3-space, albeit with nonlocal correlations.

This, then, leads to the central question of how to deal with the
nonlocality problem in a relativistic setting. We here show that a
basic argument discussing the allegedly paradox time ordering of events
in EPR-type two-particle experiments falls short of taking into account
the contextuality of the experimental setup. Consequently, we then
discuss under which circumstances (i.e.\ physical premises) superluminal
information transfer (but not signaling) may be compatible with a
Lorentz-invariant theory. Finally, we argue that the impossibility
of superluminal signaling -- despite the presence of superluminal
information transfer -- is not the result of some sort of conspiracy
(á la ``Nature likes to hide''), but the consequence of the impossibility
to exactly reproduce in repeated experimental runs a state's preparation,
or of the no-cloning theorem, respectively.%
\begin{lyxgreyedout}
\global\long\def\VEC#1{\mathbf{#1}}
\global\long\def\d{\,\mathrm{d}}
\global\long\def\e{{\rm e}}
\global\long\def\meant#1{\left<#1\right>}
\global\long\def\meanx#1{\overline{#1}}
\global\long\def\mpbracket{\ensuremath{\genfrac{}{}{0pt}{1}{-}{\scriptstyle (\kern-1pt +\kern-1pt )}}}
\global\long\def\pmbracket{\ensuremath{\genfrac{}{}{0pt}{1}{+}{\scriptstyle (\kern-1pt -\kern-1pt )}}}
\global\long\def\p{\partial}
\end{lyxgreyedout}

\end{abstract}
\maketitle

\section{Emergent Quantum Mechanics:\emph{ }realism without\protect \\
pre-determination\label{sec:Introduction}}

The term ``Emergent Quantum Mechanics'' (EmQM) has been used in
the literature for several years by now, albeit with different meanings
regarding the word ``emergence''. As a major option, the term refers
to the possibility that quantum theory might be a (very good) approximation
to some ``deeper level theory''. To some, though, EmQM just stands
for quantum theory as a special case for a particular set of parameters
of a more encompassing theory. The term would thus refer to the\emph{
emergence of a theory}. However, the meaning of EmQM may also be more
specific in that it refers to \emph{physical emergence}, i.e., to
the modeling of quantum systems as \emph{emergent systems}. It is
the latter option that our group has dealt with throughout the last
couple of years.

There is, however, a communication problem in getting the relevant
ideas across, mainly because the quantum physics and the self-organization/emergence
communities, respectively, hardly communicate with each other. Specifically,
the problem of accepting physical emergence as a possibility within
the quantum physics community seems to be the rather exotic looking
theme of \emph{top-down causation} (next to bottom-up causation).
Despite the fact that there are numerous examples in hydrodynamics,
self-organizing systems, etc., for top-down causation, most quantum
physicists seem unaffected by the possibility of this ``unusual''
type of causality, although the usual understanding of causality is
apparently insufficient for a description and understanding of quantum
processes. To give just one example, consider the Rayleigh--Bénard
cells of hydrodynamics. There, one witnesses microscopic random movement
that spontaneously becomes ordered on a macroscopic level. The top-down
causality is manifest in that emergent particle trajectories depend
crucially on the boundary conditions of the system.

Certainly, this classical example of top-down causation is neither
``weird'' (as quantum mechanics is claimed to be), nor ``surreal''
(as the trajectories' behaviors in the convection cells might be called,
did one not have a perfectly rational explanation for it). Rather,
this example must by necessity be covered by any definition of realism
that one would claim as being generally applicable to any physical
system. However, in the recent literature, definitions of realism
have been proposed in the context of quantum foundations that would
deny the above-mentioned examples of top-down causation to be considered
``real''. In other words, the phenomena of Rayleigh--Bénard cells
in particular, but also all other processes of self-organizing, or
emergent, systems, are actually counter-examples to recent definitions
of ``realism'' in the quantum foundations literature, such as the
following: ``all measurement outcomes are determined by pre-existing
properties of particles independent of the measurement (realism).''~\citep{Groblacher.2007experimental}

This is now to be contrasted with our proposal for a physical EmQM,
i.e., \emph{realism without pre-determination}. Instead of pre-determination,
in the systems of interest we consider the case of co-evolution, i.e.,
permanently updated co-determination, with essential influences on
the microphysics by changing boundary conditions, or measurement arrangements,
respectively. With this perspective, it becomes clear that via importing
ill-defined concepts of ``realism'' into the debate about Bell's
theorems, all sorts of wrong conclusions become possible. After all,
in this scenario ``realism'' effectively just becomes a ``red herring'',
with the task of killing all proposals of thus understood ``realistic''
hidden variable theories -- an exercise that in the end is rather
fruitless. Looking back on the development of quantum foundations
throughout the last decades, one actually may get the impression that
this kind of strategy to exclude realism in this way has kept a lot
of people busy. A historian of science might be well advised to consider
a study on ``A Brief History of Red Herrings'', or the like -- this
might turn into quite a voluminous book.

To give just one more example of such a red herring, not of recent
times, but of half a century ago, consider the following passage from
Richard Feynman's famous description of electrons passing a double
slit: 

``We now make a few remarks on a suggestion that has sometimes been
made to try to avoid the description we have given {[}i.e., of the
double slit experiment with electrons{]}: `Perhaps the electron has
some kind of internal works -- some inner variables -- that we do
not yet know about. Perhaps that is why we cannot predict what will
happen. If we could look more closely at the electron, we could be
able to tell where it would end up.' So far as we know, that is impossible.
We would still be in difficulty. Suppose we were to assume that inside
the electron there is some kind of machinery that determines where
it is going to end up. That machine must also determine which hole
it is going to go through on its way. But we must not forget that
what is inside the electron should not be dependent on what \emph{we}
do, and in particular upon whether we open or close one of the holes. 

So, if an electron, before it starts, has already made up its mind
(a) which hole it is going to use, and (b) where it is going to land,
we should find $P_{1}$ for those electrons that have chosen hole~\emph{1},
$P_{2}$ for those that have chosen \emph{2}, and necessarily the
sum $P_{1}+P_{2}$ for those that arrive through the two holes. There
seems to be no way around this. But we have verified experimentally
that this is not the case. And no one has figured a way out of this
puzzle. So at the present time we must limit ourselves to computing
probabilities. We say `at the present time,' but we suspect very strongly
that it is something that will be with us forever -- that it is impossible
to beat that puzzle -- that this is the way nature really \emph{is}.''~\citep{Feynman.1966vol3}

It is interesting to see how strongly Feynman insists on his conclusion
about ``the way nature really \emph{is}'', although the logic his
argument is based on depends on the unquestioned assumption that the
electron's behavior is pre-determined, and independent of the measurement.
This assumption, however, is in stark contrast to an alternative scenario
which is proposed in our EmQM, i.e., that the electron does not propagate
in a completely empty space, but is embedded in a ``medium''. The
latter we identify as the vacuum's zero-point field, which can mediate
information about the boundary conditions as given by the source and
the measurement apparatus. The zero-point field is also the decisive
agent in describing the electron's behavior in stochastic electrodynamics~\citep{Cetto.2014emerging-quantum},
with which our ansatz shares some characteristics, and a prototype
to illustrate wave-particle duality via particle-medium interactions
is given by the famous experiments with oil droplets ``walking''
on an oil bath~\citep{Couder.2005,Bush.2015pilot-wave,Batelaan.2016momentum}.

\begin{figure}[!tbh]
\begin{centering}
\includegraphics[clip,width=0.75\columnwidth]{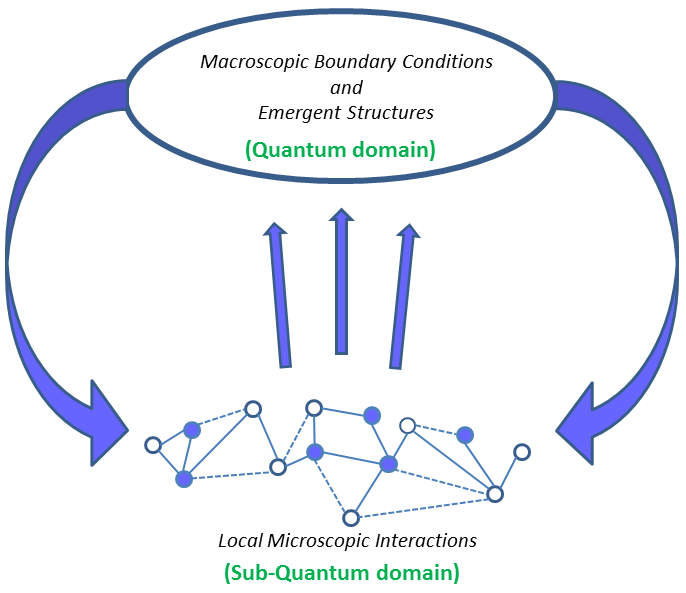}
\par\end{centering}

\centering{}\caption{{\small{}Scheme of relational causality: mutual (bottom-up and top-down)
processes at the same time.}\label{fig:interf-1}}
\end{figure}

So, there is one essential characteristic that radically contrasts
particle behavior in our EmQM to that insinuated by above-quoted definition
of ``realistic'' hidden variable theories: Instead of pre-determined
velocities of the latter, the particles in our EmQM exhibit emergent
velocities stemming from the constant interplay of the particle forward
velocity at a particular instant in time with the wave-like embedding
surroundings of the zero-point field. Said constant interplay, or
mutual influencing, of particle and wave dynamics (which has its classical
analogy in Couder's oil droplets, or ``walkers''), is a manifestation
of what we call \emph{relational causality}: the interlocking of bottom-up
and top-down causalities. Assuming local microscopic interactions
in a sub-quantum domain, these form -- together with the macroscopic
boundary conditions -- emergent structures in the quantum domain which
may exhibit spontaneous nonlocal order. In turn, the thus created
emergent structures affect the local microscopic interactions in a
top-down manner, which thus closes the causal circle relating vastly
different spatial scales at the same time (Fig.~\ref{fig:interf-1}).

\begin{figure}[!tbh]
\centering{}\includegraphics[width=1\textwidth]{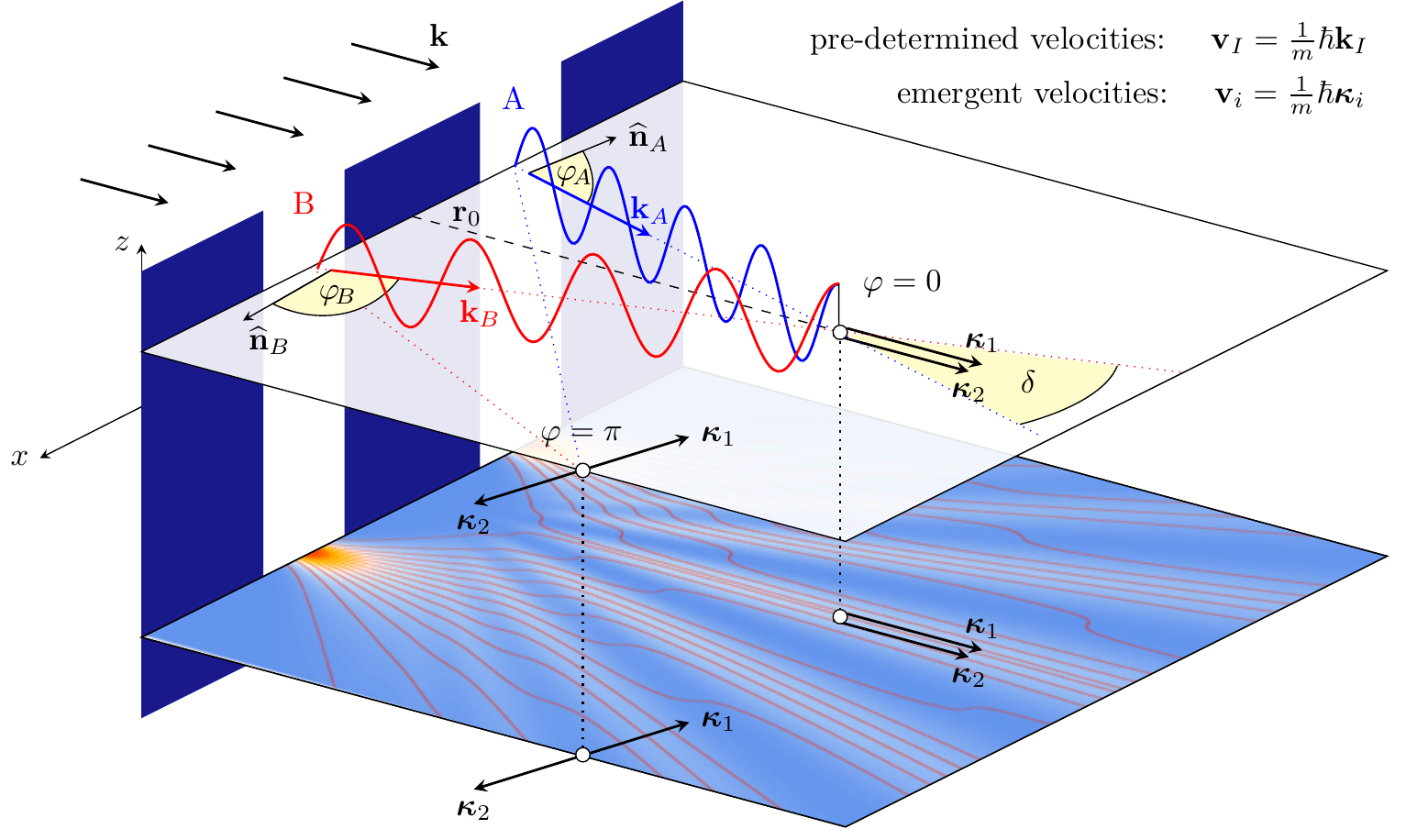}\caption{{\small{}Scheme of interference at a double slit. Considering an incoming
beam of electrons with wave number $\mathbf{k}$ impinging on a wall
with two slits, two beams with wave numbers $\mathbf{k}_{A}$ and
$\mathbf{k}_{B}$, respectively, are created, which one may denote
as ``pre-determined'' quantities, resulting also in pre-determined
velocities $\mathbf{v}_{I}=\frac{1}{m}\hbar\mathbf{k}_{I}\textrm{,\,\emph{I}=\ensuremath{A}\,or\,\ensuremath{B}.}$
Recent definitions of ``realism'' in the quantum foundations literature
would now imply that }\emph{\small{}any}{\small{} realistic hidden
variable theory just has these pre-determined velocities at its disposal
for modeling double-slit interference. However, if one considers that
the electrons are not moving in empty space, but in an undulatory
environment created by the ubiquitous zero-point field ``filling''
the whole experimental setup, a very different picture emerges. For
then one has to combine all the velocities/momenta at a given point
in space and time in order to compute the resulting, or ``emergent'',
velocity/momentum field $\mathbf{v}_{i}=\frac{1}{m}\hbar\boldsymbol{\kappa}_{i},\,\emph{i}=\ensuremath{1}\mathrm{\,or\,}\ensuremath{2}$.
The relevant contributions differing from the particle's forward momentum
$m\mathbf{v}$ originate from the so-called osmotic (or diffusive)
momentum field $m\mathbf{u}$. Thus it becomes possible to model double
slit interference in microscopic detail, as can be seen from the lower
plane where intensity distributions and average trajectories are shown.
Local momentum conservation is thereby guaranteed as soon as one takes
both the co-evolving forward and the osmotic velocity fields into
account.}\label{fig:interf-2}}
\end{figure}

We have applied the concept of relational causality to the situation
of double-slit interference (Fig.~\ref{fig:interf-2}). Considering
an incoming beam of, say, electrons with wave number $\mathbf{k}$
impinging on a wall with two slits, two beams with wave numbers $\mathbf{k}_{A}$
and $\mathbf{k}_{B}$, respectively, are created, which one may denote
as ``pre-determined'' quantities, resulting also in pre-determined
velocities $\mathbf{v}_{I}=\frac{1}{m}\hbar\mathbf{k}_{I}\textrm{, \emph{I}=\ensuremath{A}\,or\,\ensuremath{B}.}$
The definition of ``realism'' (but also Feynman's dictum) that we
criticized above would now imply that any realistic hidden variable
theory just has these pre-determined velocities at its disposal for
modeling double-slit interference. This, however, would constitute
a very naive form of realism which, to our knowledge, nobody in the
quantum physics community supports (\ldots{} thus making it a classical
red herring)! 

However, if one considers that the electrons are not moving in empty
space, but in an undulatory environment created by the ubiquitous
zero-point field ``filling'' the whole experimental setup, a very
different picture emerges. For then one has to combine all the velocities/momenta
at a given point in space and time in order to compute the resulting,
or ``emergent'', velocity/momentum field $\mathbf{v}_{i}=\frac{1}{m}\hbar\boldsymbol{\kappa}_{i}\textrm{, \emph{i}=\ensuremath{1}\,or\,\ensuremath{2}}$
(Fig.~\ref{fig:interf-2}), where $i$ is a bookkeeping index not
necessarily related to the particle coming from a particular slit~\citep{Fussy.2014multislit}.
The relevant contributions other than the particle's forward momentum
$m\mathbf{v}$ originate from the so-called osmotic (or diffusive)
momentum $m\mathbf{u}$. The latter is well known from Nelson's stochastic
theory~\citep{Nelson.1966derivation}, but its identical form has
been derived by one of us from an assumed sub-quantum nonequilibrium
thermodynamics~\citep{Groessing.2008vacuum,Groessing.2009origin}.
Introducing the osmotic momentum in a sub-quantum hidden variable
theory constitutes a decisively concrete step beyond the much older,
but only rather general proposal by one of us of ``quantum systems
as `order-out-of-chaos' phenomena''~\citep{Grossing.1989quantum}.
For now it becomes possible to model double slit interference in more
detail, with momentum conservation guaranteed as soon as one takes
both the co-evolving forward and the osmotic velocity fields into
account~\citep{Groessing.2010emergence,Fussy.2014multislit}. This
constitutes, {\small{}among others}, a viable causal model with its
implied violation of what is called ``causal parameter independence''.
The latter would state that in EPR-type scenarios Alice's measurement
outcomes would not depend on Bob's measurement settings. However,
as local changes of boundary conditions such as the settings of an
apparatus nonlocally affect the whole system, our relationally causal
model does describe said dependence and is therefore not excluded
by recent no-go principles for certain causal hidden variable theories~\citep{Ringbauer.2016experimental}.
Not only that, our model also provides an understanding and deeper-level
explanation of the microphysical, causal processes involved, i.e.,
of the ``guiding law''~\citep{Groessing.2015implications}, which
also happens to be identical with the central postulate of the de
Broglie--Bohm theory. 

\section{Identity of the emergent kinematics of \emph{$N$} bouncers in real
$3$-dimensional space with the configuration-space version of de
Broglie--Bohm theory for \emph{$N$} particles\label{sec:config}}

As in our model the ``particle'' is actually a bouncer in a fluctuating
wave-like environment, i.e.~analogously to the Couder's bouncers,
one does have some (e.g.\ Gaussian) distribution, with its center
following the Ehrenfest trajectory in the free case, but one also
has a diffusion to the right and to the left of the mean path which
is just due to that stochastic bouncing. Thus the total velocity field
of our bouncer in its fluctuating environment is given by the sum
of the forward velocity $\VEC v$ and the respective diffusive (or
``osmotic'') velocities $\VEC u_{\mathrm{L}}$ and $\VEC u_{\mathrm{R}}$
to the left and the right. As for any direction $\alpha$ the diffusion
velocity $\VEC u_{\mathrm{\alpha}}=D\frac{\nabla_{\alpha}P}{P},$$\;\alpha=L\textrm{ or \ensuremath{R,}}$
does not necessarily fall off with the distance, one has long effective
tails of the distributions which contribute to the nonlocal nature
of the interference phenomena~\citep{Groessing.2013dice}. In sum,
one has three, distinct velocity (or current) channels per slit in
an $n$-slit system. 

We have previously shown~\citep{Fussy.2014multislit,Groessing.2014relational}
how one can derive the Bohmian guidance formula from our bouncer/walker
model. Introducing classical wave amplitudes $R(\VEC w_{i})$ and
generalized velocity field vectors $\VEC w_{i}$, which stand for
either a forward velocity $\VEC v_{i}$ or a diffusive velocity $\VEC u_{i}$
in the direction transversal to $\VEC v_{i}$, we account for the
phase-dependent amplitude contributions of the total system's wave
field projected on one channel's amplitude $R(\VEC w_{i})$ at the
point $(\VEC x,t)$ in the following way: We define a \emph{conditional
probability density} $P(\VEC w_{i})$ as the local wave intensity
$P(\VEC w_{i})$ in one channel (i.e.~$\VEC w_{i}$) upon the condition
that the totality of the superposing waves is given by the ``rest''
of the $3n-1$ channels (recalling that there are 3 velocity channels
per slit). The expression for $P(\VEC w_{i})$ represents conditions
which we describe as ``relational causality'': any change in the
local intensity affects the total field, and \emph{vice versa}, any
change in the total field affects the local one. In an $n$-slit system,
we thus obtain for the conditional probability densities and the corresponding
currents, respectively, i.e.\ for each channel component $\mathit{i}$,
\begin{align}
P(\VEC w_{i}) & =R(\VEC w_{i})\VEC{\hat{w}}_{i}\cdot{\displaystyle \sum_{j=1}^{3n}}\VEC{\hat{w}}_{j}R(\VEC w_{j})\label{eq:Proj-1}\\
\VEC J\mathrm{(}\VEC w_{i}\mathrm{)} & =\VEC w_{i}P(\VEC w_{i}),\qquad i=1,\ldots,3n,
\end{align}
with
\begin{equation}
\cos\varphi_{i,j}:=\VEC{\hat{w}}_{i}\cdot\VEC{\hat{w}}_{j}.
\end{equation}
Consequently, the total intensity and current of our field read as
\begin{align}
P_{\mathrm{tot}}= & {\displaystyle \sum_{i=1}^{3n}}P(\VEC w_{i})=\left({\displaystyle \sum_{i=1}^{3n}}\VEC{\hat{w}}_{i}R(\VEC w_{i})\right)^{2}\label{eq:Ptot6-1}\\
\VEC J_{\mathrm{tot}}= & \sum_{i=1}^{3n}\VEC J(\VEC w_{i})={\displaystyle \sum_{i=1}^{3n}}\VEC w_{i}P(\VEC w_{i}),\label{eq:Jtot6-1}
\end{align}
 leading to the \textit{emergent total velocity field}
\begin{equation}
\VEC v_{\mathrm{tot}}=\frac{\VEC J_{\mathrm{tot}}}{P_{\mathrm{tot}}}=\frac{{\displaystyle \sum_{i=1}^{3n}}\VEC w_{i}P(\VEC w_{i})}{{\displaystyle \sum_{i=1}^{3n}}P(\VEC w_{i})}\,.\label{eq:vtot_fin-1}
\end{equation}

In~\citep{Fussy.2014multislit,Groessing.2014relational} we have
shown with the example of $n=2$, i.e.\ a double slit system, that
Eq.~(\ref{eq:vtot_fin-1}) can equivalently be written in the form
\begin{equation}
\VEC v_{\mathrm{tot}}=\frac{R_{1}^{2}\VEC v_{\mathrm{1}}+R_{2}^{2}\VEC v_{\mathrm{2}}+R_{1}R_{2}\left(\VEC v_{\mathrm{1}}+\VEC v_{2}\right)\cos\varphi+R_{1}R_{2}\left(\VEC u_{1}-\VEC u_{2}\right)\sin\varphi}{R_{1}^{2}+R_{2}^{2}+2R_{1}R_{2}\cos\varphi}\,.\label{eq:vtot-1}
\end{equation}

The trajectories or streamlines, respectively, are obtained in the
usual way by integration. As first shown in~\citep{Groessing.2012doubleslit},
by re-inserting the expressions for convective and diffusive velocities,
respectively, i.e.\ $\VEC v_{i}=\frac{\nabla S_{i}}{m}$, $\VEC u_{i}=-\frac{\hbar}{m}$$\frac{\nabla R_{i}}{R_{i}}$,
one immediately identifies Eq.~(\ref{eq:vtot-1}) with the Bohmian
guidance formula. Naturally, employing the Madelung transformation
for each path $j$ ($j=1$ or $2$), 
\begin{equation}
\psi_{j}=R_{j}\e^{\mathrm{i}S_{j}/\hbar},\label{eq:3.14-1}
\end{equation}
and thus $P_{j}=R_{j}^{2}=|\psi_{j}|^{2}=\psi_{j}^{*}\psi_{j}$, with
$\varphi=(S_{1}-S_{2})/\hbar$, and recalling the usual trigonometric
identities such as $\cos\varphi=\frac{1}{2}\left(\e^{\mathrm{i}\varphi}+\e^{-\mathrm{i}\varphi}\right)$,
one can rewrite the total average current immediately in the usual
quantum mechanical form as 
\begin{equation}
\begin{array}{rl}
{\displaystyle \mathbf{J}_{{\rm tot}}} & =P_{{\rm tot}}\mathbf{v}_{{\rm tot}}\\[3ex]
 & ={\displaystyle (\psi_{1}+\psi_{2})^{*}(\psi_{1}+\psi_{2})\frac{1}{2}\left[\frac{1}{m}\left(-\mathrm{i}\hbar\frac{\nabla(\psi_{1}+\psi_{2})}{(\psi_{1}+\psi_{2})}\right)+\frac{1}{m}\left(\mathrm{i}\hbar\frac{\nabla(\psi_{1}+\psi_{2})^{*}}{(\psi_{1}+\psi_{2})^{*}}\right)\right]}\\[3ex]
 & ={\displaystyle -\frac{\mathrm{i}\hbar}{2m}\left[\Psi^{*}\nabla\Psi-\Psi\nabla\Psi^{*}\right]={\displaystyle \frac{1}{m}{\rm Re}\left\{ \Psi^{*}(-\mathrm{i}\hbar\nabla)\Psi\right\} ,}}
\end{array}\label{eq:3.18-1}
\end{equation}
where $P_{{\rm tot}}=|\psi_{1}+\psi_{2}|^{2}=:|\Psi|^{2}$.

Eq.~(\ref{eq:vtot_fin-1}) has been derived for one particle in an
$n$-slit system. However, it is straightforward to extend this derivation
to the many-particle case~\citep{Groessing.2015implications}. \textsl{Therefore,
what looks like the necessity in the de Broglie--Bohm theory to superpose
wave functions in configuration space, can equally be obtained by
superpositions of all relational amplitude configurations of waves
in real 3-dimensional space. }\textsl{\emph{The central ingredient
for this to be possible is to consider the }}\textsl{emergence}\textsl{\emph{
of the velocity field from the interplay of the totality of all of
the system's velocity channels. }}

It can also be shown that the average force acting on a particle in
our model is the same as the Bohmian quantum force. For, due to the
identity of our emerging velocity field with the guidance formula,
and because they essentially differ only via the notations due to
different forms of bookkeeping, their respective time derivatives
must also be identical. Thus, from Eq.~(\ref{eq:vtot_fin-1}) one
obtains the particle acceleration field (using a one-particle scenario
for simplicity) in an $n$-slit system as
\begin{align}
\mathbf{a}_{\mathrm{tot}}\left(t\right) & =\frac{\d\mathbf{v}_{{\rm tot}}}{\d t}=\frac{\d}{\d t}\left(\frac{{\displaystyle \sum_{i=1}^{3n}}\VEC w_{i}P(\VEC w_{i})}{{\displaystyle \sum_{i=1}^{3n}}P(\VEC w_{i})}\right)\nonumber \\
 & =\frac{1}{\left({\displaystyle \sum_{i=1}^{3n}}P(\VEC w_{i})\right)^{2}}\left\{ \vphantom{\frac{{\displaystyle \sum^{3n}}}{{\displaystyle \sum^{3n}}}}\sum_{i=1}^{3n}\left[P(\VEC w_{i})\frac{\d\VEC w_{i}}{\d t}+\VEC w_{i}\frac{\d P(\VEC w_{i})}{\d t}\right]\left({\displaystyle \sum_{i=1}^{3n}}P(\VEC w_{i})\right)\right.\label{eq:3.1}\\
 & \qquad\qquad\qquad\qquad\qquad\qquad\left.-\left({\displaystyle \sum_{i=1}^{3n}}\VEC w_{i}P(\VEC w_{i})\right)\left({\displaystyle \sum_{i=1}^{3n}\frac{\d P(\VEC w_{i})}{\d t}}\right)\vphantom{\frac{{\displaystyle \sum^{3n}}}{{\displaystyle \sum^{3n}}}}\right\} .\nonumber 
\end{align}
Note in particular that~(\ref{eq:3.1}) typically becomes infinite
for regions $\left(\mathbf{x},t\right)$ where $P_{\mathrm{tot}}={\displaystyle \sum_{i=1}^{3n}}P(\VEC w_{i})\rightarrow0$,
in accordance with the Bohmian picture.

From~(\ref{eq:3.1}) we see that even the acceleration of one particle
in an $n$-slit system is a highly complex affair, as it nonlocally
depends on all other accelerations and temporal changes in the probability
densities across the whole experimental setup! In other words, this
force is truly emergent, resulting from a huge amount of bouncer-medium
interactions, both locally and nonlocally. This now leads to the central
question of how to deal with the nonlocality problem in a relativistic
setting. We shall discuss this problem by considering EPR-type experiments.

\section{Nonlocality and relativity: A case for Lorentz invariant superluminal
information transfer\label{sec:relativity}}

\begin{figure}[!tbh]
\begin{centering}
\includegraphics[clip,width=1\columnwidth]{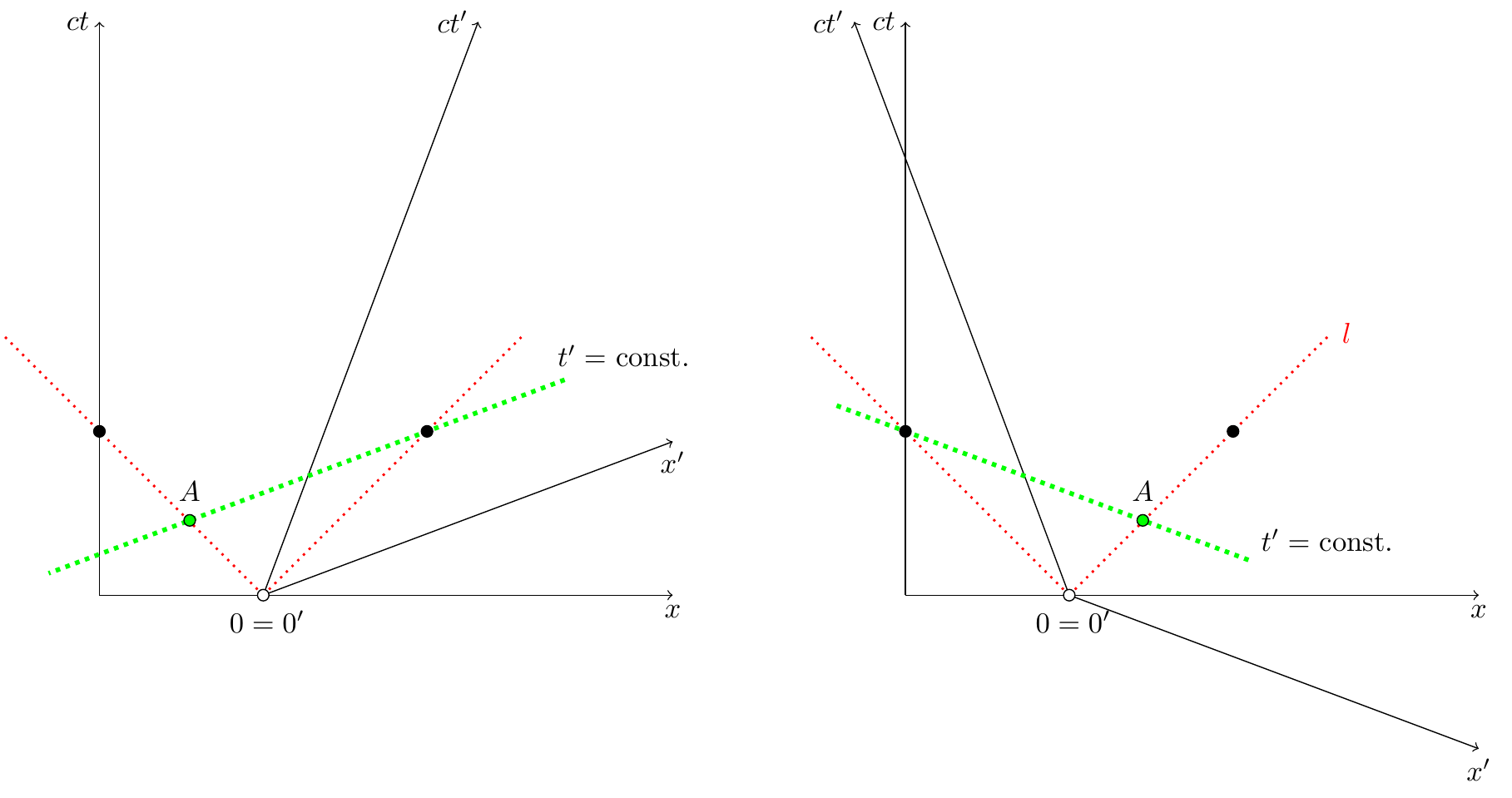}
\par\end{centering}

\begin{centering}
\vspace{-8mm}
\hfill{}(a)\hfill{}\hfill{}(b)\hfill{}
\par\end{centering}

\centering{}\caption{{\small{}(a) An EPR-correlated particle pair is emitted at source
}\emph{\small{}0}{\small{} and later simultaneously registered by
detectors (symbolized by black circles) in the laboratory rest frame.
However, an observer in a reference frame moving to the right, would
see the right-hand particle registered at a time }\emph{\small{}t'}{\small{}
at which the left-hand particle is located at spacetime point }\emph{\small{}A}{\small{}.
Thus, a ``state jump'' would occur at }\emph{\small{}A}{\small{}
before the particle is registered. (b) Same as in (a), but with observer
in a reference frame moving to the left. The latter will see the left-hand
particle registered at time }\emph{\small{}t'}{\small{} at which the
right-hand particle is located at spacetime point }\emph{\small{}A}{\small{},
so that a ``state jump'' would occur at }\emph{\small{}A}{\small{}
before the particle is registered. Taken together, (a) and (b) point
at a conflict as to which order of events ``really'' is happening
in a realistic world view.}\label{fig:superlu-3-1}}
\end{figure}

As is well known, in an EPR-type experiment, where two photons are
emitted in opposite directions from one spin-zero state, two different
observers would form mutually inconsistent pictures of reality (Fig.~\ref{fig:superlu-3-1}).

However, Maudlin has correctly pointed out that ``Relativity also
reveals some of the apparent contradictions between frames to be merely
matters of equivocation.''~\citep{Maudlin.2011quantum} He explains:
``The unprimed (here: l.h.s.) frame says that the right-hand photon
is detected before the left while the primed (here: r.h.s.) frame
has it the other way around. How could they both be right? In this
case the answer is clear: they are simply talking about different
things. The unprimed frame notes precedence in its \emph{t}-coordinate,
which we might call `time', while the primed frame is concerned with
precedence in its own \emph{t}-coordinate, which we could call `primetime'.
There is no more contradiction between saying that the right detection
event precedes the left in time but follows it in primetime than there
is in saying that Idaho precedes New Jersey in geographical area but
follows it in terms of population.''~\citep{Maudlin.2011quantum} 

\begin{figure}[!tbh]
\begin{centering}
\includegraphics[clip,width=1\columnwidth]{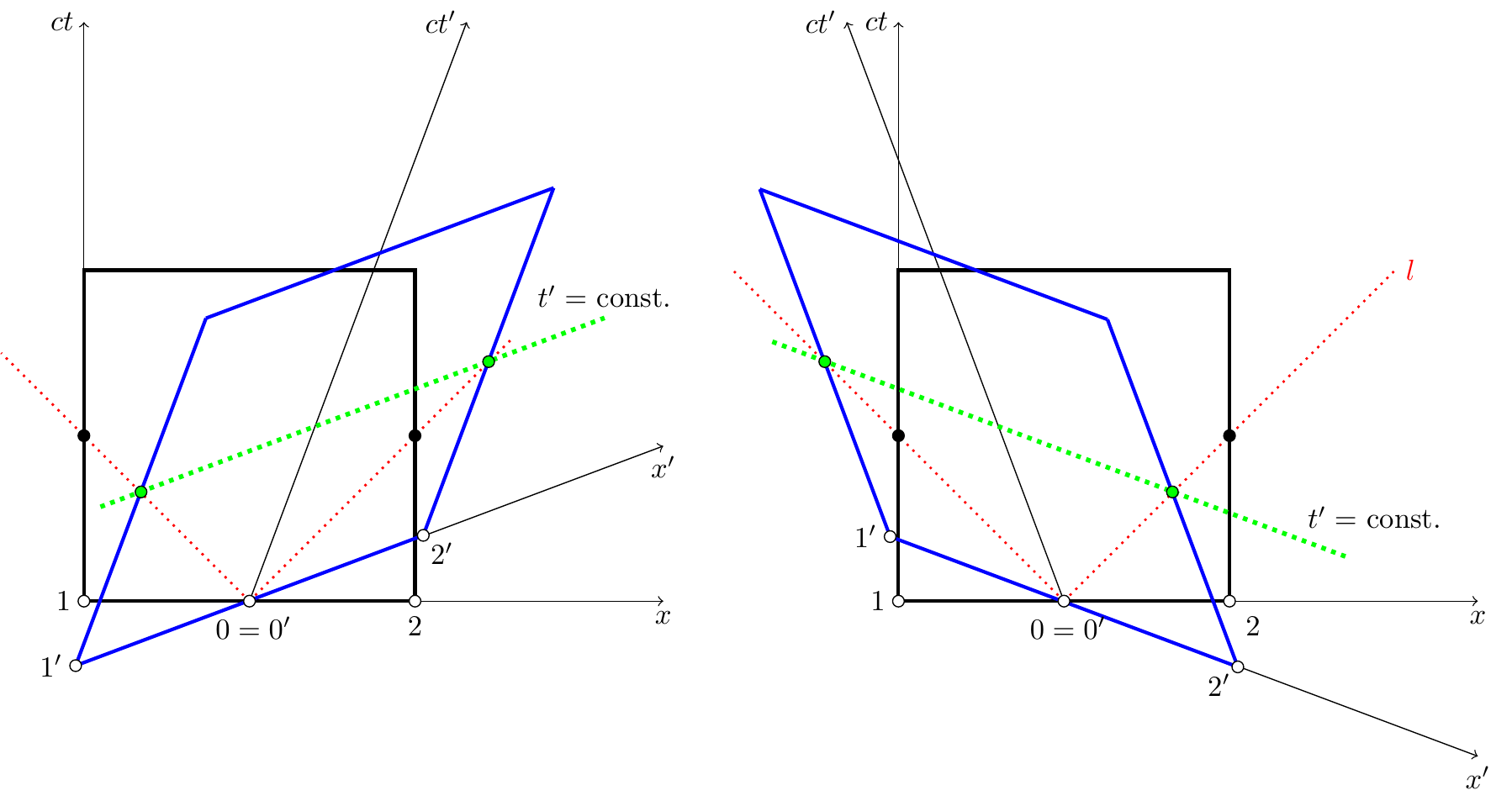}
\par\end{centering}

\begin{centering}
\vspace{-8mm}
\hfill{}(a)\hfill{}\hfill{}(b)\hfill{}
\par\end{centering}

\centering{}\caption{{\small{}As Lorentz transformations correspond to simple rotations
in spacetime diagrams, the two situations of Fig.~\ref{fig:superlu-3-1}
just correspond to two different viewpoints. However, what is crucially
important to be able to show this, is that one has to take into account
the whole experimental arrangement including the source }\emph{\small{}0}{\small{}
and the detectors }\emph{\small{}1}{\small{} and }\emph{\small{}2}{\small{},
in each reference frame. Then, even nonlocal correlations (symbolized
by the green line) are part of a relativistically invariant description.
Moreover, this renders obsolete the question of which of the options
(a) and (b) was the ``true'' time ordering.}\label{fig:superlu-3-2}}
\end{figure}

So, the riddle posed in Fig.~\ref{fig:superlu-3-1} is solved by
realizing that there does not exist the one ``true time ordering''.
All that matters is the Lorentz invariance of the theory. Paraphrasing
Maudlin, we add ``space'' and ``primespace'' in the scenario,
just to obtain frames of ``spacetime'' and ``prime-spacetime''.
Then, our two different viewpoints are two versions of a Lorentz invariant
behavior in space and time, which can be transformed into each other
via simple rotation, as required by the Lorentz transformations (Fig.~\ref{fig:superlu-3-2}).

Again, we agree with Maudlin as to the consequences of this insight.
He argues that many would agree that ``Relativity prohibits something
from going faster than light: 

Matter or energy cannot be transported faster than light. 

Signals cannot be sent faster than light. 

Causal processes cannot propagate faster than light. 

Information cannot be transmitted faster than light.'' 

But, alternatively, one just needs to require that theories must be
Lorentz invariant: ``This requirement is compatible with the violation
of every one of the prohibitions listed above.''~\citep{Maudlin.2011quantum}
However, in order to clearly see the implications of the required
Lorentz invariance, one must appreciate that the whole experimental
arrangement has to be taken into account, i.e., including the source
\emph{0} and the detectors \emph{1} and \emph{2}, in each reference
frame. Then, even nonlocal correlations (symbolized by the green line
in Fig.~\ref{fig:superlu-3-2}) are part of a relativistically invariant
description.

\begin{figure}[!tbh]
\begin{centering}
\includegraphics[clip,width=0.5\columnwidth]{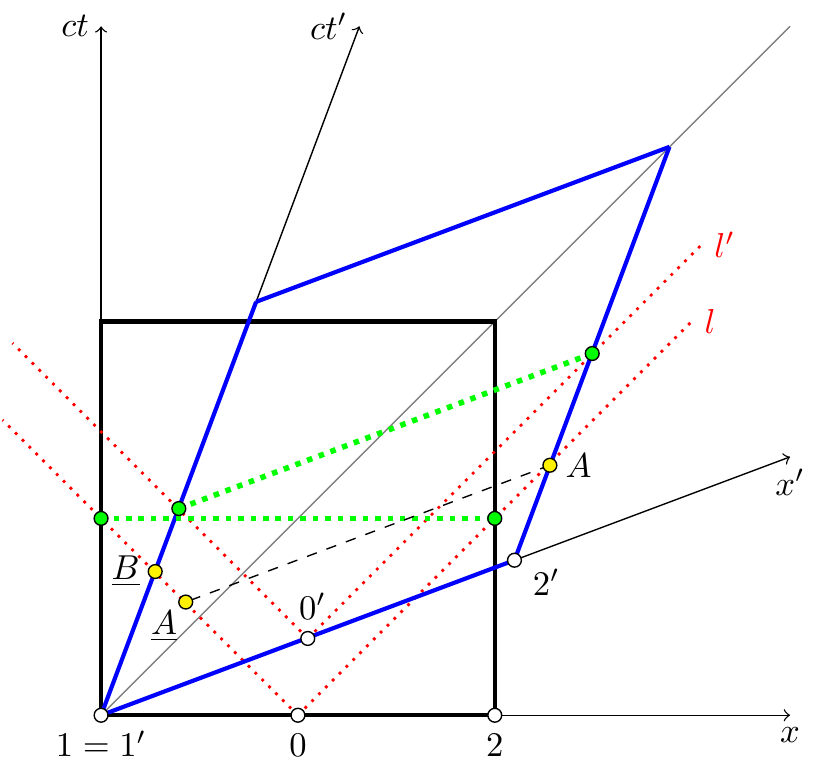}
\par\end{centering}

\centering{}\caption{{\small{}In the laboratory rest frame, the EPR-correlated particle
pair is emitted at source }\emph{\small{}0}{\small{} and later simultaneously
registered by detectors (symbolized by green circles). If, under the
wrong assumption, the moving observer would register at }\emph{\small{}A}{\small{}
a photon that was emitted at }\emph{\small{}0}{\small{}, then the
``state jump'' of the left-hand photon would occur at $\underline{A}$,
i.e., before its arrival at the detector at $\underline{B}$. However,
since the preparation of an entangled photon pair requires the preparation
of the whole experimental setup, and not just of the source, the world
lines of all elements of the apparatus must be considered as forming
one unseparable whole. Therefore, the photon pairs can be emitted
from source 0' only at times $t_{0}^{'}\geq0$. Consequently, also
in the moving frame will the photons arrive simultaneously at the
respective detectors, and Lorentz invariance is again established.}\label{fig:superlu-3-3}}
\end{figure}

An especially interesting scenario is given for the case that the
two frames meet such that the spacetime points at the origin coincide,
i.e.~\emph{1}=\emph{1'} (Fig.~\ref{fig:superlu-3-3}). In the resting
laboratory frame, the experiment is arranged along an area between
points \emph{1} and \emph{2}, with the source \emph{0} emitting photon
pairs at time $t_{0}\geq0$. Imagine now that to each element of the
experimental setup is attached a traffic light that shows ``red''
when the preparation is not yet completed, and ``green'' when it
is. In the rest frame, the totality of all traffic lights showing
``green'' will occur at some first instance, i.e., at some time
$t_{0}=\textrm{const}$. Now, what would a moving observer see? The
answer is given by Rindler's ``wave of simultaneity'': whereas the
green lights will light up in a ``flash'' which in the rest frame
occurs simultaneously at the time $t_{0}=0$, in the moving observer's
frame they will light up simultaneously at some time $t_{0}^{'}=0$.
(For a more detailed description, with arguments involving a ``rigged
Hilbert space'', see~\citep{Groessing.2000qcbook}.) This has the
strict consequence that the photon pairs can be emitted from source
\emph{0'} only at times $t_{0}^{'}\geq0$. And this is also the source
of misleading arguments in the literature. For, although the lightcone
is defined uniquely in all reference frames, this is true only as
to its spreading in spacetime due to the universality of the vacuum
speed of light. However, the \emph{timing} of the emission of the
photon pairs must in general be different for two observers moving
with relative velocities to each other. In other words, one has to
consider the relativity of simultaneity of the whole experimental
setup including the source and the extension of the apparatus. Note
also that if the moving observer would register at \emph{A} a photon
that was emitted at \emph{0}, we would have the same kind of dilemma
as in Figs.~\ref{fig:superlu-3-1} (a) and (b). For then the ``state
jump'' of the left-hand photon would occur at $\underline{A}$, i.e.,
before its arrival at the detector $\underline{B}$. However, if one
correctly describes the source \emph{0'} of the particle pair in the
moving observer's rest frame, then the detectors at points \emph{1'}
and \emph{2'}, respectively, would always simultaneously register
the corresponding photons. Thus, by relating the axis of equitemporality
to the whole extension of an experimental apparatus in spacetime,
which is different for each reference frame, one avoids the error
of attributing some ``idealistic'' time to the system (like, e.g.,
$t'=\textrm{const}$ in Figs.~\ref{fig:superlu-3-1} (a) and (b)).

We have thus shown that even nonlocal correlations can be part of
a relativistically invariant description of events in spacetime. The
conditions for Lorentz-invariant superluminal information transfer
are essentially given by the evolution in spacetime of the whole,
encompassing system (i.e., quantum system plus macroscopic apparatus,
including source and detectors). A crucial question then arises as
to interventions in such a system, which lead to the phenomenon of
\emph{dynamical nonlocality~}\citep{Tollaksen.2010quantum}. In the
above-mentioned paper by one of us~\citep{Grossing.1989quantum},
a type of experiment called ``late choice experiment'' was proposed
that we then believed would provide an effect with dramatic consequences.
The question was clarified for us only fairly recently in the paper
by Tollaksen \emph{et al}.~\citep{Tollaksen.2010quantum}, with the
consequence that although the effect exists its measurable consequences
are not dramatic at all. We shall here only refer to the variant that
Tollaksen \emph{et al}.\ discuss, for it is the relevant one. Essentially,
the authors ask what happens in a realistic scenario (where one electron
goes through just one of two slits present) if at the very moment
the particle passes the slit, the other slit is being closed (or opened,
in case it was closed before). The authors show within the Heisenberg
picture that the opening or closing of a slit results in the nonlocal
transfer of (what they call ``modular'') momentum. (Our group has
shown that this effect can also be demonstrated when using the Schrödinger
picture~\citep{Groessing.2013dice}.) 

However, there is an in-principle uncontrollability of that momentum
transfer: Tollaksen \emph{et al}.~\citep{Tollaksen.2010quantum}
speak of ``complete uncertainty'' in this regard. This means that
here one has the case of a nonlocal transfer of information (i.e.,
from a slit to the particle), which, however, cannot be used for signaling:
due to the necessary uncertainty of the location of the particle before
the intervention, the nonlocal momentum transfer just shifts the particle
within the wave packet, figuratively speaking, so that its detection
at some location cannot in principle give any indication of whether
or not that information transfer has happened. 

A corollary to this concerns an implication of the no-cloning theorem.
Opening or closing of a slit amounts to the preparation of a new state
with a certain phase relative to the one associated with the path
of the particle through the other slit. Considering many runs with
such an intervention highlights an important feature of the system:
it cannot be controlled. In other words, the identical phase can \emph{never}
be reproduced, which means that the no-cloning theorem is here ultimately
responsible for the impossibility of superluminal signaling. (For,
if one could control the phase during intervention at a slit, one
could do this massively in parallel, and thus on average steer the
electron to a desired position.) 

Thus, both the consistencies of relativity and of quantum theory are
confirmed, despite the superluminal information transfer involved.
In other words, complete uncertainty makes it possible to have \emph{nonlocal
information transfer without superluminal signaling}. The latter is
exactly the option shown in~\citep{Walleczek.2016nonlocal} to remain
valid for viable nonlocal hidden variable theories.

\begin{turnpage}

\end{turnpage}

\begin{ruledtabular}
\end{ruledtabular}

\begin{turnpage}

\end{turnpage}

\appendix

\begin{acknowledgments}
We thank Jan Walleczek for many enlightening discussions, and the
Fetzer Franklin Fund of the John E. Fetzer Memorial Trust for partial
support of the current work.
\end{acknowledgments}

\providecommand{\href}[2]{#2}\begingroup\raggedright\endgroup

\end{document}